\documentstyle[11pt]{article}
\def\p{\partial} \def\D{{\cal D}}  \def\bD{{\bar {\cal D}}}
  \def\bW{{\overline W}}  
\def\e{\epsilon} \def\g{\gamma} \def\n{\nabla} \def\d{\delta}
\def\r{\rho} \def\s{\sigma}   
\def\c{\chi} \def\t{\vartheta}  \def\bt{{\bar \vartheta}}
\def\b{\beta} \def\a{\alpha} \def\l{\lambda}  \def\f{\varphi}
\def\da{{\dot \alpha}} \def\db{{\dot \beta}}  \def\dg{{\dot \gamma}}
\def\dd{{\dot \delta}} \def\hs{harmonic space } \def\L{{\cal L}}
\def\sd{self-dual } \def\ssd{super self-dual } 
\def\sdy{self-duality } \def\ssdy{super self-duality }
\def\sym{super Yang-Mills } \def\ym{Yang-Mills }

\def\half{{1\over 2}}
\def\der#1{{\partial \over \partial #1}}
\def\be{\begin{equation}} \def\r#1{(\ref{#1})}
\def\ee{\end{equation}}
\def\ba{\begin{array}{rll}} \def\ea{\end{array}}
\font\sqi=cmssq8
\def\R{\rm I\kern-1.45pt\rm R}
\def\C{\kern2pt {\hbox{\sqi I}}\kern-4.2pt\rm C}
\def\op{\oplus} \def\om{\ominus}
\begin{document}
\rightline{hep-th/9410147}
\vskip1.1truein
\leftline{{\large {\bf Four dimensional integrable theories
\footnote{Talk by V. Ogievetsky at the G\"ursey Memorial Conference I,
Istanbul, June 1994}
}}}
\vskip 1.0 truecm
\leftline{Ch. Devchand  and  V. Ogievetsky}
\vskip.1truein
\leftline{Joint Institute for Nuclear Research
141980 Dubna, Russia}
\vskip .65truein
\noindent
{\bf Abstract.}
There exist many four dimensional integrable theories. They include
self-dual gauge and gravity theories, all their extended supersymmetric
generalisations, as well the full (non-self-dual) N=3 super Yang-Mills
equations. We review the harmonic space formulation of the twistor
transform for these theories which yields a method of producing explicit
connections and metrics. This formulation uses the concept of harmonic space
analyticity which is closely related to that of quaternionic analyticity.
\vskip.65truein
\noindent{\bf 1. Introduction} \vskip 8pt
Many Lorentz invariant four dimensional exactly solvable nonlinear theories
are known. The most remarkable of these are those admitting the Penrose-Ward
twistor transform \cite{tw}, which may be thought of as an analogue of the
transformation to action-angle variables for hamiltonian dynamical systems,
in the sense that it involves a transformation to variables in which the
dynamics is trivial, reducing the problem to that of inverting the
transformation. Further, the solution methods for many lower-dimensional
completely integrable systems, like the inverse scattering transform for
the KdV equation, may be thought of as reductions of the twistor transform
\cite{Ward}, so the prospect has arisen, of a unification of the various
existing methods of solving two dimensional systems as different
manifestations of the twistor transform for \sd \ym (SDYM).

The twistor transform, which takes its most dramatic form
in its application to the solution of the self-dual \ym and Einstein
equations has been found to have a remarkably clear realisation in the
language of `harmonic spaces' (\cite{h}-\cite{gio}). In fact harmonic or
twistor spaces admit supersymmetrisation, yielding a remarkably simple
supersymmetrisation of the SDYM and Einstein equations, which is much more
straightforward, and moreover independent of the N-extension (where N is the
number of independent supersymmetries), than the supersymmetrisations of the
corresponding full non-self-dual theories, for which the supersymmetrisation
for each extension N has to be considered anew.
All N-extended supersymmetric theories may therefore
be treated on an equal footing \cite{ssd,sdsg}. Moreover, for the \sd \sym
theories, there exists a remarkable `matreoshka'-like nested structure
\cite{ssd} in which
the $N=0$ solution data may be dressed-up to higher N solution data in
a basically algebraic fashion using solutions of first-order equations.

The list of four-dimensional theories (which may equally well be
considered to be in complexified space or in real spaces of Euclidean (4,0)
or Kleinian (2,2) signature) amenable to the twistor transform is
therefore quite large and includes
\begin{itemize}
\item  Self-dual \ym (SDYM) equations, for any semisimple gauge group.
\item  All N-extended ($N=1,..,4$) supersymmetrisations of the latter.
\item  Self-dual Einstein equations, with or without cosmological constant.
\item  All N-extended Poincar\'e and conformal \sd supergravities.
\item  The full (i.e. non-\sd) N=3 super-\ym theory (even in the
       Minkowskian (3,1) signature).
\end{itemize}
In this talk, we shall describe the harmonic space versions of the twistor
transform for all the above theories.
The crucial feature allowing the applicability of the twistor transform
to field theories is the possibility of presenting the equations of motion
in the form of algebraic constraints amongst the components of some
curvature tensor, the paradigmatic example being the \ym \sdy equations.
In particular, the constraints take the form
$$ [\D_{\a i},\D_{\b j}] = \e_{\a\b} F_{ij} , $$
where $\a,\b$ are spinor indices of some group having skew-symmetric
invariant $\e_{\a\b}$, $i,j$ are some other indices or labels, and $F_{ij}$
are the non-zero curvatures representing the obstruction to Frobenius'
integrability. Twistor or harmonic space is an auxiliary space in which the
curvature is zero in some `analytic' subspaces, allowing the use of
`Frobenius variables' to reduce the system. In the harmonic space setting a
transformation to such variables converts the system to a set of
Cauchy-Riemann-like (CR) equations, thereby reducing the problem to that of
reconstructing the original variables from the `analytic' data
(satisfying these CR equations).
The crucial idea of harmonic space analyticity is closely related to
the concepts of quaternionic and Fueter analyticity, to which Feza G\"ursey,
whom we all loved so dearly, devoted so much attention. It is therefore
especially appropriate to present these ideas at this meeting dedicated to
his memory. In fact it was precisely in Feza's last paper (with
V. Ogievetsky and M. Evans)\cite{GOE} that the intimate relation between
quaternionic and harmonic space analyticities was clarified. That
paper was completed shortly after Feza's untimely death and we feel
it appropriate to quote the dedication to Feza contained in its manuscript,
which Physical Review refused to include in the published version.

``Feza G\"ursey, a fine human being and outstanding physicist,
passed away on April 13, 1992. He is a coauthor of the present paper,
which is one of a series of his works devoted to quaternionic
aspects of four-dimensional field theories, a field in which he
was a pioneer.
Feza enthusiastically participated in the writing of this paper, even
as he fought the disease to which he finally succumbed.
Sadly, he did not live long enough to approve the paper's final version,
and so bears no responsibility for whatever shortcomings it
may possess. It was a great joy and privilege to work with
Feza, and to benefit from his fertile mind and keen intelligence.
The experience of working with him and the wonderful
personality of Feza G\"ursey will abide forever in the memories of the
two other authors. ''
\goodbreak \noindent
{\bf 2. From 2D {\em complex} to 4D {\em quaternionic} analyticity}\vskip 8pt
In two-dimensional Euclidean space the two {\it real} coordinates
may be quite naturally combined into {\it a single complex number}
$x^\mu=\{x^1, x^2\} \rightarrow z=x^1 + i x^2 $ and the most general
conformal coordinate transformation in two dimensions is the {\it analytic}
transformation \be z'=f(z), \quad  \bar z' = \bar f(\bar z). \label{0}\ee
In virtue of the Cauchy-Riemann condition,
${\partial\over\partial {\bar z}} f(z)= 0$, its d'Alembertian vanishes,
${\partial\over\partial z}{\partial\over\partial{\bar z}}f(z)=0.$

Similarly naturally, four dimensional coordinates may be combined into
a quaternion. In spinor notation we have
\be
x^\mu \rightarrow  q= x^{\alpha \dot\alpha}=
\pmatrix{  x^0 - ix^3 & -ix^1 - x^2 \cr
                       -ix^1 + x^2 & x^0 + ix^3 } =
x^0 + e_a x^a         \label{2}
\ee
where the Pauli matrices represent the algebra of the quaternionic units,
$e_a=-i \sigma_a$
\be
e_a e_b =- \delta_{ab} + \epsilon_{abc} e_c.     \label{3}
\ee

Analytic transformations (\ref{0}) are fundamental to 2D-conformal
field theories. Feza G\"ursey often wondered whether there exist
4D theories in which some form of quaternionic analyticity  plays a
correspondingly crucial r\^ole \cite{oldgu}, \cite{gur1}.
However, the notion of quaternionic analyticity is rather delicate
and there are several possible forms, some of which being too restrictive
to be applicable to field theories. For instance, the straightforward
generalisation of the Cauchy-Riemann condition
\be {\partial\over\partial {\bar q}} f=
\frac{1}{2} \left ({\partial\over\partial{x^0}}+
\frac{1}{3} e_a {\partial\over\partial {x^a}}\right ) f=0      \label{4}
\ee
where ${\partial\over\partial{\bar q}}$ satisfies
${\partial\over\partial{\bar q}} q = 0 $ and
${\partial\over\partial{\bar q}} \bar q = 1 $
is well known  (see e.g. \cite{sud}) to allow only a linear solution
$f= a + qb $, with constant
quaternions $a$ and $b$, because of the noncommutativity of quaternions.

{\em Fueter quaternion analyticity} \cite{fuet,gj}, however, is less
restrictive. This defines an analytic
function of a quaternion $q$, as a Weierstrass-like series
\be f(q) = \sum a_n q^n,          \label{F} \ee
where the coefficients $a_n$ are real or complex numbers (or quaternions,
but multiplying $q^n$ on only one side, e.g. left as in (\ref{F})).
Such a function obeys a Cauchy-Riemann-like condition, of the {\it third
order} in derivatives and is therefore in general not a harmonic function
($\Box f(q)\neq 0 $), although it is bi-harmonic ($ \Box^2 f(q) =0 $).
Moreover, it is not invariant under SO(4) rotations \cite{gj}.

In  self-dual  and  $N=2$ supersymmetric theories, however,  manifolds of
quaternionic character namely quaternionic-K\"ahler and hyper-K\"ahler
manifolds naturally arise \cite{gib}.
In these theories hyper-K\"ahler and quaternionic structures are
related to yet another notion of analyticity, namely harmonic-space
analyticity, which we shall explain.
\vskip 8pt
\noindent{\bf 3. Harmonic space}\vskip 8pt
Harmonic space \cite{h} is essentially an enlargement of four dimensional
space-time, which may be thought of in terms of the coset space
${\mbox{ Poincar\'e group}\over \mbox{ Lorentz group}}$, to coset space
${\mbox{ Poincar\'e group}\over SU(2) \times U(1)} =
{\mbox{ Poincar\'e group}\over \mbox {Lorentz group}}
\times {SU(2)\over U(1)}$
(for the case of signature (4,0)). This space has additional coordinates
parametrising the two-sphere $S^2 = {SU(2)\over U(1)}$.
Of course, one could choose polar ($\theta,\phi$) or stereographic
($z,\bar z$) coordinates to describe this sphere. However, it is in practice
very useful to use a more abstract parametrisation using two fundamental
representations of the SU(2) algebra, $u^\pm_\da$ (where $\da$ is an SU(2)
spinor index and $\pm$ denote U(1) charges), which are just spin
$\half$ spherical harmonics of $S^2$, defined up to the U(1) equivalence
$u^{\pm}_\da \sim e^{\pm\g} u^\pm_\da\; ;\g \in \C$ and satisfying
the equations $ \e_{\da\db} u^{+\da} u^{-\db} = 1 $.
The further hermiticity condition $ u^-_\da=\overline{u^{+\da}} $
yields two independent real variables. In the complexified setting,
however, $u^+_\da $ and $u^-_\da$ are independent and an appropriate
equivalence relation holds \cite{GOE}.
\vskip 8pt
\goodbreak
\noindent{\bf 4. Self-dual \ym } \vskip 7pt
The usual \sdy condition for the \ym field strength
\be F_{\mu\nu} = \half\e_{\mu\nu\rho\s} F_{\rho\s}\  , \label{sd}\ee
basically says that the (0,1) part of the gauge field vanishes. This is
better expressed in terms of 2-spinor notation in the form:
$ f_{\da\db} =\ 0  $ which is equivalent to the statement that the field
strengths curvature only contains the (1,0) Lorentz representation, i.e.
\be [\D_{\a\da} ~,~\D_{\b\db}] =\  \e_{\da\db} f_{\a\b} .\label{sd2}\ee
Now multiplying \r{sd2} by the two commuting spinors $u^{+\da}, u^{+\db}$,
one can compactly represent it as the vanishing of a curvature
\be [\n^+_\a , \n^+_\b ]  =  0\  ,\label{zc}\ee
where $ \n^+_\a \equiv u^{+\da}\n_{\a\da}$, with linear system
\be \n^+_\a \f =\ 0\ .\label{ls}\ee
This is precisely the Belavin-Zakharov-Ward linear system for SDYM.  Now
the $ u^{+\da}$ are actually harmonics \cite{h} on $S^2$ and it is better
to consider these equations in an auxiliary space with
coordinates $\{ x^{\pm \a} \equiv x^{\a\da}u^\pm_\da ~ ,~ u^\pm_\da ~~;\;
u^{+\da}u^-_\da=1 \} $, where the harmonics are defined up to a $U(1)$
phase, and gauge covariant derivatives in this harmonic
space are
\be    \n^+_\a = \p^+_\a  + A^+_\a  = \der{x^{-\a}} + A^+_\a  .\label{cov}\ee
In this space \r{zc} is actually not equivalent to the \sdy conditions.
We also need
\be [ D^{++} , \n^+_\a ]=  0\ ,\label{lin}\ee
where $D^{++}$ is a harmonic space derivative which acts on
negatively-charged harmonics to yield their
positively-charged counterparts, i.e. $D^{++} u^-_\da =  u^+_\da $,
whereas $ D^{++} u^+_\a  = 0$. This operator, in a fixed
parametrisation, has also been considered by Newman (e.g. \cite{newman}).
In ordinary x-space, when the
harmonics are treated as parameters, the condition \r{lin} is actually
incorporated in the definition of $\n^+_\a$ as a {\em linear} combination
of the covariant derivatives.  The system (\ref{zc},\ref{lin}) is now
{\em equivalent} to SDYM and has been considered by many authors, e.g.
\cite{GOE,ym}; the equivalence holding in spaces of
signature (4,0) or (2,2), or in complexified space. In this regard, we
should note that for real spaces, our understanding is completely clear
for the Euclidean signature. For the (2,2) signature, the situation is
richer and more intricate due to the noncompact nature of the rotation group
and our present considerations concern only those signature (2,2)
configurations which may be obtained by Wick rotation of (4,0) configurations.

Now, in \r{lin} the covariant derivative \r{cov} has pure-gauge form
\be   \n^+_\a = \p^+_\a  + \f \p^+_\a \f^{-1} .\label{pg}\ee and $D^{++}$
is `short' i.e. has no connection. This choice of frame (the `central'
frame) is actually inherited from the four-dimensional x-space and is not
the most natural one for harmonic space. We may however change coordinates
to a basis in which  $\n^+_\a$ is `short' and  $D^{++}$ is `long' (i.e.
acquires a Lie-algebra-valued connection) instead. Namely,
\be\ba  \n^+_\a =&\ \p^+_\a \\[2mm]
    \D^{++} =&\ D^{++} + V^{++} ,\ea\ee  a change of frame tantamount
to a gauge transformation by the `bridge' $\f$ in \r{ls}. In this basis
(the `analytic' frame) the SDYM system (\ref{zc},\ref{lin}) remarkably takes
the form of a Cauchy-Riemann (CR) condition
\be \der{ x^{-\a} }  V^{++} =\ 0 \label{cr}\ee
expressing independence of half the x-coordinates. In virtue of passing to
this basis the nonlinear SDYM equations \r{sd} are in a sense trivialised:
Any `analytic' (i.e. satisfying \r{cr}) function
$V^{++} = V^{++}(x^{+\a}, u^\pm)$ corresponds to some \sd gauge
potential. From any such  $V^{++}$, by solving the {\em linear} equation
\be D^{++}\f =\  \f V^{++} \label{b}\ee for the bridge $\f$, a \sd vector
potential may be recovered from the harmonic expansion:
\be \f \p^+_\a \f^{-1} =\  u^{+\da} A_{\a\da} \label{con};\ee
the linearity in the harmonics $u^{+\da}$ being guaranteed by \r{lin}.

Solving \r{b} for  an {\em arbitrary} analytic gauge algebra valued
function $V^{++}$ yields the {\em general} local \sd solution. This
correspondence between \sd gauge potentials and holomorphic prepotentials
$V^{++}$ is a convenient tool for the explicit construction of local
solutions of the \sdy equations.

Furthermore, in the analytic subspace of harmonic space (with coordinates
$\{x^{+\a}, u^\pm_\da \}$), there exists an especially simple presentation
of the infinite-dimensional symmetry group acting on solutions of the \sdy
equations. It is the (apparently trivial) transformation
$V^{++} \rightarrow V^{++'} =  g^{++},$ where $g^{++}$ depends in an
arbitrary way on $V^{++}$ and its derivatives as well as on the analytic
coordinates themselves, modulo gauge transformations
$V^{++} \rightarrow e^{-\l}(V^{++} + D^{++})e^\l $, where $\l$ is also
an arbitrary analytic function.
\vskip 8pt
\goodbreak
\noindent{\bf 5. Supersymmetric \sd\ym theories}
\vskip 8pt
Yang-Mills theories can be supersymmetrised to couple successively lower
spin fields to the vector field.
Since extended \sym theories are massless theories, the components are
classified by helicity and we have the following representation content in
theories up to N=3:
\be\begin{array}{lccccccccc}
   &helicity:&    1       &  \half   & 0 &  -\half
   &  \half   & 0   &  -\half   & -1                                 \\[2mm]
                       &  N=0   &  f_{\a\b}  &          &   &
   &          &     &           &  f_{\da\db}                        \cr
                       &  N=1   &  f_{\a\b}  & \l_\a    &   &
   &          &     &  \l_\da   &  f_{\da\db}                        \cr
                       &  N=2   &  f_{\a\b}  & \l^i_\a  & \bW  &
  &          &  W  & \l_{\da i} &  f_{\da\db}                       \cr
                       &  N=3   &  f_{\a\b}  & \l^i_\a  & W_i  & \c_\da
  & \c_\a    & W^i & \l_{\da i} &  f_{\da\db}                       \cr
\end{array}\label{tri}\ee
In real Minkowski space fields in the left and right triangles are related
by CPT conjugation but in complexified space or in a space with signature
(4,0) or (2,2), we may set fields in one of the triangles to zero
without affecting fields in the other triangle. If we set all the fields
in the right (left) triangle to zero, the equations of motion reduce to
the super (anti-) \sdy equations. For instance, the \sdy equations
for the N=3 theory take the form
\be\ba
\e^{\b\g} \D_{\g\db} f_{\a\b} =& 0\\[7pt]
 \e^{\g\b} \D_{\g\db} \l_{\b}^i =& 0 \\[7pt]
\e^{\dg\da} \D_{\a\dg} \c_{\da } =& - [\l_{\a}^k , W_k]  \\[7pt]
 \D_{\a\db}\D^{\a\db} W_i  =& \half \e_{ijk}\{\l^{\a j}, \l_\a^k\}
       .\ea\label{n3}\ee
We see that the spin 1 source current actually factorises into parts from
the two triangles, so it manifestly vanishes for \ssd solutions.
The first equation in \r{n3} is just the Bianchi identity for \sd
field-strengths. So apart from the \sdy condition \r{sd}, we have
one equation for  zero-modes of the covariant Dirac operator in the
background of a \sd vector potential (having \r{sd} as integrability
condition) and two further non-linear equations. However, any given
\sd vector potential actually {\em determines} the general (local)
solution of the rest of the equations. This is the most striking
consequence of the matreoshka phenomenon: the N=0 core determining
the properties of the higher-N theories. Another consequence is
is that many conserved currents identically vanish in the \ssd sector.
For instance, since \sdy always implies the {\em source-free} second order
\ym equations, the spin 1 source current vanishes for the entire matreoshka.
Further, the usual \ym stress tensor clearly vanishes for \sd fields:
$$ T_{\alpha{\dot \alpha},\beta{\dot \beta}}
\equiv\ f_{{\dot \alpha}{\dot \beta}} f_{\alpha\beta} =\ 0\    ;$$
and as a consequence of this, once
one has put on further layers of the matreoshka, the supercurrents
generating supersymmetry transformations, which contain the stress tensor
as well as its superpartners also identically vanish for \ssd fields.

In N-independent form, \r{n3} can be conveniently written as the
following super curvature constraints in chiral superspace:
\be\ba
\{ \bD^{i}_{\da} , \bD^j_{\db} \}& = & \e_{\da\db} W^{ij} \\[7pt]
     \{\D_{\a i}, \D_{\b j}  \}& =&  0 = [ \D_{\a i} , \n_{\a\b} ]  \\[7pt]
\{ \D_{\a j} , \bD^i_\db \}& =&  2 \d^i_j \n_{\a\db} \  .\ea\label{ssd}\ee
Having expressed the \ssdy  equations in this form, the supersymmetrisation
of the harmonic-twistor construction is straightforward.
In harmonic superspaces with coordinates
$$\{ x^{\pm \a} \equiv u^\pm_\db x^{\a\db},\
  \bt^\pm_i \equiv u^\pm_\da \bt^{\da}_i ,\ \t^{\a i} ,\ u^\pm_\da \},$$
these take the form
\be\ba
     \{\D_{\a i}, \D_{\b j}  \} = & 0 & =  \{ \bD^{+i} , \bD^{+j} \}  \\[7pt]
        [\n^+_\a , \n^+_\b ]  =  & 0 & =  [ \bD^{+i} , \n^+_\a ] \\[7pt]
\{ \D_{\a j} , \bD^{+i} \}& = & 2 \d^i_j \n^+_\a  \\[7pt]
[ \D_{\a i} , \n^+_\b  ]  & = &  0, \ea\label{ssdh}\ee
where the gauge covariant derivatives are given by
\be
 \D_{\a i} = D_{\a i} + A_{\a i} ,\quad
         \bD^{+i} = u^{+\da} \bD^{i}_{\da}= \bar D^{+i} + \bar A^{+i} ,\quad
   \n^+_\a = u^{+\da} \n_{\a\da} = \p^+_\a  + A^+_\a  \  ,\ee
and satisfy the equations
\be [ D^{++} , \D_{\a i} ]  = [ D^{++} , \bD^{+i}] = [ D^{++} , \n^+_\a ]
=  0\  .\label{ssdcr}\ee
The equations (\ref{ssdh},\ref{ssdcr}) are {\em equivalent} to \r{ssd}
and \r{ssdh} are consistency conditions for the following system of
linear equations
\be\ba \D_{\a i} \f & = 0  \cr
             \bD^{+i} \f & = 0 \cr
             \n^+_\a \f &  = 0 , \ea\ee
This system is extremely redundant, $\f$ allowing the following
transformation under the gauge group
\be  \f \rightarrow
e^{-\tau(x^{\a\da}, \bt_i^\da \t^{\a i})} \f
                        e^{\l(x^{+\a}, \bt_i^+ , u^\pm_\da)}\  ,\ee
where $\tau$ and $\l$ are arbitrary functions of the variables shown,
without affecting the constraints \r{ssdh}. These constraints therefore
allow an economic choice of chiral-analytic basis in which the bridge
$\phi$ and the prepotential $V^{++}$ depend only on
{\em positively $U(1)$-charged, barred} Grassmann variables,
viz. $\bt^+_i$, being independent of $\t^{i\a}$ and $\bt^-_i$.
In this basis, $\f$ too is independent of $\t^{i\a}$ and $\bt^-_i$; its
non-analyticity manifesting itself in its dependence on $x^{-\a}$.
Moreover, consistently with the commutation relations \r{ssdh},
the covariant spinor derivatives take the form
$ \D_{\a i} = \der{\t^ {\a i}},\bD^i = 2 \t^{\a i} \n^+_\a .$ The
\ssdy conditions (\ref{ssdh},\ref{ssdcr}) are therefore equivalent to
the {\em same} system of equations as the N=0 SDYM equations, viz.
(\ref{ls},\ref{lin}), except that now
$\f$ and $A^+_\a$ are chiral superfields depending on
$\{ x^{\pm \a} ,\bt^+_i, u^\pm_\da \}$ \cite{ssd}. As for the N=0 case,
we may express this system in the form
of analyticity conditions for the harmonic space connection superfield
$V^{++}$:
\be \der{ x^{-\a} }  V^{++}(x^{+\a}, \bt^{+i} ,u^\pm_\da)  =\ 0
\label{scr}.\ee
and the bridge $\f$ to the central basis may be found by solving \r{b}.
Fields solving \r{n3} may then be obtained by inserting solutions $\f$
of \r{b} into the expression
\be \f \p^+_\a \f^{-1} =\  u^{+\da} A_{\a\da}(x^{\a\da}, \bt_i^\da)  ,\ee
(the left side being guaranteed to be linear in $u^+$), and expanding the
superfield vector potential on the right to obtain the component multiplet
satisfying \r{n3} thus:
\be  A_{\a\db}(x, \bt) = A_{\a\db}(x) + \bt_{\db i} \l^i_\a(x)
                + \e^{ijk} \bt_{\da j} \bt^\da_i \n_{\a\db} W_k(x)
      + \e^{ijk} \bt_{\da i} \bt^\da_j \bt^\dg_k \n_{\a\dg}\c_\db(x)\ .\ee
It is remarkable that \ssdy implies the absence of higher-order terms in
$\bt$. In fact any N=0
solution completely and recursively determines its higher-N extensions
\cite{ssd}.

The most general infinite-dimensional group of transformations of
super-self-dual solutions acquires a transparent form in the analytic
harmonic superspace with coordinates $\{ x^{+\a},\bt^{+i},u^{\pm}_\da\}$.
As for the $N=0$ case, it is given by the transformation
\be V^{++} \rightarrow V^{++'}  =
g^{++}(V^{++}, x^{+\a}, \bt^{+i}, u^{\pm}_{\da}),\label{g}\ee
where $g^{++}$ is an arbitrary doubly $U(1)$-charged analytic algebra-valued
functional, modulo gauge transformations
$V^{++} \rightarrow e^{-\l}(V^{++} + D^{++})e^\l $, where $\l$ is also
an arbitrary analytic function.
This group has an interesting subgroup of transformations
\be V^{++} \rightarrow V^{++'} = V^{++}(x^{+'}, \bt^{+'}, u')
,\label{dif}\ee
induced by diffeomorphisms of the analytic harmonic superspace
\be x^{+\a'} = x^{+\a'}(x^+,\bt^+, u),\quad
 \bt^{+i'} = \bt^{+i'}(x^+,\bt^+, u),\quad
 u' = u'(x^+,\bt^+, u).  \label{dif1} \ee
\vskip 8pt
\goodbreak
\noindent{\bf 6. N=3 (non-self-dual) \sym theory}
\vskip 8pt
As we have seen, the spin 1 source currents of all \ssd theories vanish
because they factorise into parts from the two triangles in \r{tri}.
It turns out that we can restore these source currents and solve the full
(i.e. non-self-dual) \sym equations by intermingling \sd
and anti-\sd holomorphic data \cite{witten}; and this works {\em exactly}
for the N=3 case. Again the crucial feature is the presentability of
the thrice-extended super Yang-Mills equations in the form of the
super-curvature constraints \cite{GSW}
\be\ba
   \{\D_{i\a} ~,~\D_{j\b}\} =&\ \e_{\a\b} W_{ij}   \\[7pt]
  \{\bD^{i}_{\da} ~,~\bD^{j}_{\db}\} =&\ \e_{\da\db} W^{ij}  \\[7pt]
  \{\D_{i\a} ~,~\bD^j_{\db}\} =&\  2\delta^j_i \n_{\a\db} ,\label{1}\ea\ee
where $\D_A \equiv \p_A +  A_A = (\n_{\a\db}, \D_{i\a}, \bD^j_{\db}),
i,j = 1,2,3, $ are gauge-covariant super-derivatives. These
constraints are purely kinematical for N=1,2 but are equivalent to
the dynamical equations for the component fields for N=3 \cite{hhls}.
Now in order to present these as zero-curvature conditions in some harmonic
space, the appearance of invariants of both simple parts of the Lorentz
group $(\e_{\a\b},\e_{\da\db})$ requires the `harmonisation' of the
entire Lorentz group. This allows the consideration of all possible
signatures, with the corresponding harmonic spaces being given by:
$$\ba
Euclidean \quad (4,0): &
{\mbox{ Poincar\'e group}\over \mbox {Lorentz group}} \times
{SU(2)\over U(1)} \times {SU(2)\over U(1)}  \cr
Lorentzian \quad (3,1):&
{\mbox{ Poincar\'e group}\over \mbox {Lorentz group}} \times
{SL(2,\C)\over L(1,\C)}\cr
Kleinian \quad (2,2): &
{\mbox{ Poincar\'e group}\over \mbox {Lorentz group}}\times
{SL(2,\R)\over SO(2)}\times{SL(2,\R)\over SO(2)} \ea$$
To thus harmonise the entire Lorentz group, we need to introduce
harmonics with both dotted and undotted indices: $u^+_\da ,u^-_\da$ and
$v^\op_\a , v^\om_\a$, satisfying the constraints
$$ u^{+\da} u_\da^{-} = 1  ~,~ v^{\oplus\a} v_\a^{\ominus} = 1 $$
and having the hermiticity condition
$\overline{u^+_\da} = v^{\om \a}$ for the Lorentzian signature.
Now in harmonic space with coordinates
$ u^{+}_{\da}, u^{-}_{\da}$ and $v^{\ominus}_{\a}, v^{\oplus}_{\a}$
and
$$  x^{\pm\oplus} =   x^{\a\da} u^{\pm}_\da  v^{\oplus}_\a ~,~
    x^{\pm\ominus} =  x^{\a\da} u^{\pm}_\da  v^{\ominus}_\a , $$
$$     \t^{i\oplus} =  \t^{i\a}  v^{\oplus}_\a ~,~
     \t^{i\ominus}=  \t^{i\a}  v^{\ominus}_\a ~,~
      \bt_i^{\pm} = \bt_i^{\da} u^{\pm}_\da ;$$
The superspace constraints \r{1} are equivalent
to the following system of equations in harmonic superspace
\be\ba
\{\bD^{+i} ~,~ \bD^{+j} \}& =&\ 0\ =\
\{ \D^{\oplus }_i ~,~ \D^{\oplus }_j \}\\[7pt]
\{ \bD^{+j} ~,~ \D^{\oplus }_i \}& =&\  2 \n^{+\oplus}   \\[7pt]
[D^{++}, \bD^{+j}]& =&\ 0 =\  [D^{++}, \D^{\oplus }_i ]\
=\  [D^{++} , \n^{+\oplus}] \\[7pt]
[D^{\oplus \oplus} , \bD^{+j}]& =&\ 0 =\ [D^{\oplus \oplus},\D^{\oplus }_i]\
=\ [D^{\oplus \oplus} ,  \n^{+\oplus}] \\[7pt]
 [ D^{++} , D^{\oplus \oplus} ]& =&\ 0 ,\label{n3cr} \ea\ee
where   $$ D^{++} =\ u^{+ \da} \der {u^{- \da}}, \quad
D^{\op\op}= v^{\op \a} \der {v^{\om \a}} .$$
Now as before, we can go to an `analytic frame' in which the covariant
derivatives $(\bD^{+j},\D^{\oplus }_i,\n^{+\oplus})$ lose their connection
parts and the derivatives $(D^{++} , D^{\oplus \oplus})$ acquire
connections $(V^{++} , V^{\oplus\oplus})$  instead. For the latter, \r{n3cr}
are just the generalised Cauchy-Riemann `analyticity' conditions
$$\ba
 {\p \over\p {\bar\t_i^-}}     V^{++} =&\ 0
                     & = {\p \over\p {\bar\t_i^-}} V^{\oplus \oplus}\\[7pt]
 {\p \over\p {\t^{\ominus i}}} V^{++} =&\ 0
                &= {\p \over\p {\t^{\ominus i}}}V^{\oplus \oplus} \\[7pt]
 {\p \over\p {x^{-\ominus}}}        V^{++} =&\  0
&   = {\p \over\p {x^{-\ominus}}} V^{\oplus\oplus}  \ea   $$
together with the zero-curvature condition
$$D^{++} V^{\oplus\oplus} - D^{\oplus\oplus} V^{++} + [V^{++} ,
V^{\oplus\oplus}] = 0, $$
which relates the two harmonic space connections $(V^{++},V^{\oplus\oplus})$.
Analytic $(V^{++},V^{\oplus\oplus})$ satisfying this relationship therefore
encode the solution of N=3 \sym theory \cite{n3ym}.
\goodbreak \vskip 8pt
\noindent{\bf 7. Self-dual gravity and supergravity}
\vskip 8pt
Analogously to \r{sd2} \sd gravity may be described by the equation
\be [\D_{\a\da} ~,~\D_{\b\db}] =\  \e_{\da\db} R_{\a\b} ,\label{sdg}\ee
where now the covariant derivative contains a vierbein as well as a
connection,
\begin{equation} {\cal D}_{\alpha \da} =
E^{\mu \db}_{\alpha \da}\partial_{\mu\db}
+ \omega_{\da\alpha}\   ,\label{sdg1}\end{equation}
so \r{sdg} is not only a curvature constraint on the components of
the connection, but also a zero-torsion condition on the vierbein. Moreover,
since the Riemann tensor has irreducible components
$$\begin{array}{rl}
R_{\alpha \beta} &\equiv
C_{(\alpha\beta\gamma\delta)} \Gamma^{\gamma\delta}
 + R_{(\alpha\beta)(\dg\dd)} \Gamma^{\dg\dd} +
{1\over 6}R \Gamma_{\alpha\beta} ,\cr
R_{\da \db} &\equiv
C_{(\da\db\dg\dd)}\Gamma^{\dg\dd} + R_{(\g\d)(\da\db)} \Gamma^{\gamma\delta}
+ {1\over 6} R \Gamma_{\da\db},\cr\end{array}$$
where $C_{(\da\db\dg\dd)} (C_{(\alpha\beta\gamma\delta)}) $ are the
(anti-) self-dual components of the Weyl tensor,
$R_{(\alpha\beta) (\dg \dd)}$  are the components of the tracefree Ricci
tensor, $R$ is the scalar curvature,
$(\Gamma^{\gamma\delta},\Gamma^{\dg\dd})$ are generators of the tangent
space gauge algebra, self-duality, i.e. the vanishing of $R_{\da\db}$
clearly implies that the curvature takes values only in one SU(2) algebra
with generators $\Gamma^{\gamma\delta}$, so we may work  in a
`self-dual  gauge' in which the connection also takes values only in
this SU(2), i.e. only this half of the
tangent space group is localised, while the second $SU(2)$
remains rigid. Restricting the holonomy group in this fashion, the
curvature part of \r{sdg} is automatic, these equations reducing to
the zero torsion conditions on the vierbein. Now, since we have to
deal with the vanishing of torsions, the \hs system equivalent to
\r{sdg} is rather different to that in the \sd \ym case. It takes the
form \be\ba
 [{\cal D}^+_\alpha, {\cal D}^+_\beta] &=&0 \\[5pt]
 [{\cal D}^{++}, {\cal D}^+_\alpha] &=& 0 \\[5pt]
 [{\cal D}^+_\alpha, {\cal D}^-_\beta] &=& 0
 \quad\hbox{(modulo  $R_{\alpha \beta}$)} \\[5pt]
[{\cal D}^{++}, {\cal D}^-_\alpha ] &=&{\cal D}^+_\alpha .\ea\ee
Now, going to Frobenius coordinates
\begin{equation} x^{\mu a} \rightarrow  x^{\mu \pm}_h
=  x^{\mu \pm}_h (x^{\mu a}u^\pm_a , u^\pm_a)\  ,\label{6}\end{equation}
in which the covariant derivative $\D^+_\a = \p^+_\a$, the partial
derivative, all the dynamics gets concentrated in the vielbeins and
connection components of
$$\D^{++}= \partial^{++} + H^{++ \mu +}\partial^-_{h \mu} +
(x^{\mu +}_h + H^{++ \mu -}) \partial^+_{h\mu}  + \omega^{++}.$$ These
may be solved for \cite{sdg} in terms of an arbitrary analytic prepotential
$\L^{+4}$ and the problem reduces to finding the explicit functions \r{6}
for any specified choice of $\L^{+4}$. Inverting the transformation \r{6} the
\sd vierbein then allows itself to be decoded. An explicit illustration of
the procedure may be found in \cite{sdg}, where the simplest monomial
choice of prepotential, $\L^{+4} = g(x_h^{1+} x_h^{2+})^2 $, where g is a
dimensionful parameter, is shown to correspond to the \sd Taub-NUT metric.
\goodbreak
Remarkably, the N-extended supersymmetric \sdy equations allow themselves
to be expressed in  chiral superspace in the same form as \r{sdg},
\begin{equation} [{\cal D}_{B \db}, {\cal D}_{A \da}] =
                  \epsilon_{\da \db} R_{AB},\label{hk}\end{equation}
except that now the indices $A,B$ are `superindices' of the superalgebra
$OSp(N|1)$. The explicit construction of the \sd super-vielbein
therefore closely follows that for the non-supersymmetric case. This
yields interesting non-trivial supersymmetrisations of hyper-K\"ahler
manifolds. In \cite{sdsg} we construct some explicit examples of
super deformations of flat space
(with curvature only in the odd directions) and of Taub-NUT space.
\vskip 8pt
\noindent{\bf 8. Open problems}
\vskip 8pt
We have discussed a large class of four dimensional integrable systems
allowing solution using the harmonic-twistor transform. Our considerations
have raised a number of interesting questions. Integrability in two
dimensions is known to imply remarkable constraints on the S-matrix
yielding factorisation into two-particle amplitudes. Whether the
integrability of the four dimensional theories discussed here has any
analogous consequences, either for these thories themselves or for their
dimensional reductions, remains an open question. This question is
especially interesting for the full N=3 \ym theory, which is known to be an
ultraviolet finite field theory. A further intriguing open question is
what class of non-\sd solutions to the usual N=0 \ym equations
can be obtained by reduction of this supersymmetric construction; and
whether the existence of two spectral parameters (corresponding to
the two sets of harmonics) yields new classes of lower dimensional
exactly solvable systems.

The remarkable conjunction of maximal supersymmetrisation, ultraviolet
finiteness, and classical integrability in the sense described here,
suggests the need to investigate the full (non-\sd) Poincar\'e and conformal
supergravity theories in this light as well. The corresponding super-twistor
construction remains an open problem.

We should like to thank the Erwin Schr\"odinger Institute, Vienna,
where this lecture was written up, for generous hospitality.

\end{document}